\documentstyle[prd,aps,doublespace,12pt]{revtex}
\begin{document}
\title{Stochastic Noise as a Source of  Decoherence in a Solid State Quantum Computer }
\author{C.J.Wellard and L.C.L. Hollenberg.}
\address{Special Research Centre for Quantum Computer Technology, School of Physics, University of Melbourne,Victoria 3010,AUSTRALIA.}
\address{email CJW : cjw@physics.unimelb.edu.au}
\maketitle
\begin{abstract}
\par We examine a stochastic noise process that has a decohering effect on the average evolution of qubits in
the quantum register of the solid state quantum computer proposed
by Kane \cite{Kane}. We consider the effects of this process on
the single qubit operations necessary to perform quantum logical
gates and derive an expression for the fidelity of these gates in
this system. We then calculate an upper bound on the level of this
stochastic noise tolerable in a workable quantum computer.
\end{abstract}
\newpage
\section{Introduction}
\par The process of computation by quantum logic, so called quantum computation, has recently been shown to be far
more powerful for solving certain classes of problem than is
classical computation \cite{Feynman,Shor,Grover}. The superiority
stems from the ability of the quantum analogue of bits, qubits, to
maintain coherence between different classical states. This allows
the quantum computer to perform computations over many classical
input states at once, giving a quantum computer an exponential
increase in speed for solving certain problems. In order to
exploit this advantage in quantum computation it is vital that the
evolution of the qubits be not only coherent, but precisely known
to the operator. In this paper we examine the evolution of qubits
under the influence of a stochastic noise process, the effect of
which is  to make the exact evolution of the system uncertain. We
thus consider an ensemble of qubits and calculate their average
evolution. We find that the effect of the noise is to produce a
decay of the average phase coherence of the qubits in the quantum
register and to depolarize qubits undergoing single qubit
operations. This ensemble decoherence manifests itself as a decay
of the fidelities of the quantum operations the qubits are
undergoing. We calculate this fidelity and use it to determine an
upper bound on the level of stochastic noise that the computer can
tolerate yet still operate successfully within the limits set by
current error correcting codes.
\section{The Kane Solid State Quantum Computer}
\par Throughout this paper we will be considering a solid state quantum computer (QC), as proposed by Kane \cite{Kane}.
In this system the qubits are simply spin ${1\over2}$ $^{31}$P
nuclei, in a silicon substrate. The system is subject to a
background magnetic field oriented in the $z$ direction, $B_z$. At low
energies the effective Hamiltonian for the nucleus-electron system
is given by \cite{Kane,Goan}
\begin{equation}
H_{n+e} = \mu_B B_z \sigma^z_e - g_n\mu_n B_z \sigma^z_n + A {\vec \sigma_e}.{\vec \sigma_n}.
\label{equation:hexact}
\end{equation}
$A = {8 \over 3} \pi \mu_B g_n \mu_n |\psi(0)|^2$ and
$|\psi(0)|^2$ is the probability density of the electron wave
function evaluated at the position of the nucleus. We can alter A
by applying a voltage to a so called ``A-gate'' situated above the
nucleus . This applied voltage shifts the electron wave function
away from the nucleus and thus reduces A. If $A<<2 B_z \mu_B$ the
electron spin states are separated in energy by a factor of
approximately $\mu_B / (g_n \mu_n) = 1633.8 $ greater than the
nuclear states. Thus the nuclei can be manipulated without
significantly altering the electron's state. We therefore consider
as our quantum computing basis a sub-space of the entire Hilbert
space spanned by the Hamiltonian Eq(\ref{equation:hexact}), namely
the states $|\downarrow 0\rangle , |\downarrow 1\rangle$. This
corresponds to the basis of nuclear states with the electron in
its ground state. The effective Hamiltonian in this sub-basis to first order in $A/(\mu_B B_z)$, up to a constant, is given by
\begin{equation}
H = B_z \gamma \sigma^z.
\label{equation:reg-nn}
\end{equation}
For convenience we have omitted the $n$ subscript on the Pauli operator as we shall do for the rest of the paper. We can tune the Lamour frequency of the nucleus via the A-gate bias; 
\begin{equation}
\gamma = -g_n \mu_B - {A_0 - \eta V \over B_z},
\end{equation}
 where $V$ is the applied A-gate voltage, $\eta = 5 \pi \times 10^7 Hz/V$ and $A_0$ is the value of $A$ when $V=0$.
\par The single qubit operations are implemented by the application of an oscillating transverse magnetic field. This
field can be brought into resonance by tuning $\gamma$ to satisfy
the resonance condition $\omega = 2 B_z \gamma / \hbar$. In this
way specific qubits can be operated on without affecting the rest
of the qubits in the register. The Hamiltonian for single qubit rotations then becomes $
H_{sqr} = H + H(t) $, where
\begin{equation}
H(t) = - B_{ac} g_n \mu_n ( \cos(\omega t + \phi)\sigma^{y}
-\sin(\omega t + \phi)\sigma^{x}).
\end{equation}
Here the phase factor $\phi$ determines
the axis of the rotation. Without loss of generality we set $\phi
=0$ and consider only $y$ rotations. We convert to a frame
rotating at this resonance frequency by transforming to the
interaction picture, and find
\begin{equation}
\tilde{H}_{sqr} = - B_{ac} g_n \mu_n  \sigma^y.
\label{equation:rf-nn}
\end{equation}
\section{Decoherence of the Quantum Register}
\par We now consider the effect of a stochastic white noise in the applied A-gate voltage, that is we write the
voltage signal
\begin{equation}
V(t) = V_0 (1 + \Delta (t)),
\label{equation:vt}
\end{equation}
where $\Delta (t)$ describes a white noise process \cite{Gardiner}. We can thus write
\begin{equation}
\Delta (t) dt = {\sqrt \lambda}dW(t),
\label{equation:lambda}
\end{equation}
where $ dW(t) $ is the Wiener increment, and ${\sqrt \lambda}$
scales the noise. We can calculate $\lambda$ by integrating
Eq(\ref{equation:vt}) over the duration of the voltage pulse, $\tau$,
to give the pulse area:
\begin{equation}
\Gamma(\tau) = V_0 \tau + \Delta \Gamma(\tau).
\end{equation}
Here $ \Delta \Gamma(\tau)$ is a Gaussian random variable with a mean of
zero and a variance $ V_0^2 \lambda \tau $. The ratio
of the rms value of the fluctuations in the pulse area to average
pulse area is given by:
\begin{equation}
{\Delta \Gamma(\tau)_{rms} \over {\bar \Gamma(\tau)}}={\sqrt{\lambda \over \tau}}.
\end{equation}
We find then, that the Hamiltonian for a qubit in the quantum
register, that is a qubit not undergoing an operation ($B_{ac}=0$), in the
presence of this noise is given by
\begin{equation}
H = B_z (\gamma  + \xi(t))\sigma^z.
\end{equation}
Here $\xi(t)$ gives the stochastic fluctuations in the Lamour
frequency of the qubit caused by the noise, it is related to the
noise in the voltage signal by
\begin{equation}
\xi(t) = {\eta \hbar V_0 \over B_z}\Delta(t).
\label{equation:xi}
\end{equation}
We define $ \xi (t) dt = {\sqrt \epsilon} dW(t) $, where
\begin{equation}
\epsilon = ({\eta \hbar V_0 \over  B_z})^2 \lambda.
\label{equation:epslam}
\end{equation}
\par Let us transform into a frame rotating at the Lamour frequency, that is transform to the interaction
picture as we have already done for the Hamiltonian
Eq(\ref{equation:rf-nn}). In this picture a qubit in a noiseless
quantum register does not evolve at all, the Hamiltonian is zero.
However if we include the white noise, the evolution is generated
by the Hamiltonian
\begin{equation}
{\tilde H} = B_z \xi (t) \sigma^z.
\end{equation}
From this we can form the Ito stochastic differential equation for the density operator in the interaction picture
\begin{eqnarray}
d{\tilde \rho} (t) &=& {B_z \over i \hbar} {\sqrt \epsilon}dW(t) [\sigma^z,{\tilde \rho} (t)] \nonumber\\
  &-& {\epsilon B_z^2 \over 2 \hbar^2} [\sigma^z,[\sigma^z,{\tilde \rho} (t)]] dt.
\end{eqnarray}
Because we do not know the precise history of the evolution, we take an average over noise histories and thus calculate
the evolution of the average density operator. Using $\langle dW(t) \rangle = 0$ we find
\begin{equation}
{d {\tilde \rho_{av}} (t)  \over dt} = - {\epsilon B_z^2 \over 2 \hbar^2} [\sigma^z,[\sigma^z,{\tilde \rho_{av}} (t) ]].
\label{equation:master_register}
\end{equation}
Master equations with this nested commutator structure have been studied extensively in the field of quantum optics
\cite{Milburn1,Milburn2,Schneider}. Let us define polarization vector by ${\tilde \rho_{av}}(t) = {1 \over 2} ({\bf 1}
+ \vec{P}(t).\vec{\sigma})$. We then consider the evolution of the average polarization vector in the quantum computation
basis. In this notation the information about coherence is contained in the $x$ and $y$ components of the polarization
vector. The magnitude of the polarization vector gives the purity of the state; if the state is completely pure then
$|\vec{P}| = 1$, in a completely mixed state the magnitude is $0$ and partially pure states have a magnitude between
these two extremes. The evolution leads to an exponential decay of the $x$ and $y$ components, indicating a loss of
phase coherence, while the $z$ component remains unchanged:
\begin{eqnarray}
{d P_z(t)  \over dt } &=& 0 ,\\ {d P_{x,y}(t)  \over dt } &=& - {
2 \epsilon B_z^2 \over \hbar} P_{x,y}(t).
\end{eqnarray}
Thus phase coherence is destroyed but population probabilities are conserved. These equations have the solution
\begin{eqnarray}
P_z(t) &=& P_z(0), \nonumber\\ P_{x,y}(t) &=& {\rm exp}[{- 2 B_z^2
\epsilon t \over \hbar^2}]P_{x,y}(0),
\end{eqnarray}
in which we can explicitly see the exponential decay of phase
coherence. We can see how this decoherence of the average density
operator effects the operation of the QC by calculating the
average fidelity of the operation. This gives the probability that
the state evolves as we expect it to if there were no noise, and
can be obtained by calculating the trace of the product of the
evolved density operator and the density operator we would expect
from a noisless evolution, in this case the zero time density
operator.
\begin{eqnarray} F &=& {\rm Tr} [ {\tilde \rho_{av}}(t)
{\tilde \rho}_{av}(0)],\nonumber\\ &=& {1 \over 2}(1 + P_z(0)^2 +
(P_x(0)^2 + P_y(0)^2){\rm exp}[{- 2 B_z^2 \epsilon t \over
\hbar^2}]). \label{equation:fidel}
\end{eqnarray}
We can see that the  fidelity depends on the initial state of the
system, specifically on how much phase coherence the initial state
possessed. If the qubit is initially in a classical state the
system cannot decohere and we get a fidelity of 1, a perfect gate.
Generally however, the initial state will be some kind of
superposition of classical states and coherence will be destroyed,
leading to a loss of fidelity. The worst case is when the initial
state is a maximum superposition, the system decoheres, and the
fidelity is given by
\begin{equation}
F = {1\over 2} (1 +{\rm exp}[{- 2 B_z^2 \epsilon t \over
\hbar^2}]). 
\label{equation:worstfidel}
\end{equation}
In this case the fidelity eventually decays to a limiting value of
${1 \over 2}$. Because the process is only phase destroying, a
measurement of $\sigma^z$, that is a projection onto the basis of
classical states will yield the same result as in the noiseless
case. In this sense we can say that the classical information has
been retained, the qubit has been converted to a classical bit.
\section{Fidelity of Single Qubit Operations}
\par To calculate the effect of the stochastic fluctuations in the A-gate voltage bias on a qubit undergoing a single
particle operation we consider the example of a rotation around the $y$-axis. The inclusion of white noise transforms
the Hamiltonian Eq(\ref{equation:rf-nn}) to
\begin{equation}
\tilde{H}_{sqr} = \xi (t) B_z \sigma^z - B_{ac} g_n \mu_n \sigma^y.
\label{equation:rf-int}
\end{equation}
We can then form the Ito stochastic differential equation
\begin{eqnarray}
d\tilde {\rho}(t) &=& -{B_{ac} g_n \mu_n \over i \hbar}  [\sigma^y,\tilde{\rho}(t)]dt \nonumber +{B_z \sqrt{\epsilon}
dW(t)\over{i \hbar}} [\sigma^z,\tilde{\rho}(t)]\nonumber\\
&-& {\epsilon B_z^2 dt \over 2 \hbar^2} [ \sigma^z,[\sigma^z,\tilde{\rho}(t)]].\end{eqnarray}
Again we average over the noise histories and find
\begin{equation}
{d{\tilde \rho_{av}}(t) \over dt}= - {B_{ac} g_n \mu_n \over i \hbar}  [\sigma^y, \tilde {\rho_{av}}(t)] - {\epsilon B_z^2 \over
2 \hbar^2} [\sigma^z ,[ \sigma^z , \tilde{\rho_{av}}(t)]].
\end{equation}
This gives a set of coupled differential equations for the polarization vector components of the average density
operator
\begin{eqnarray}
{d P_x(t)\over dt} &=& {-  2 B_z^2  \epsilon \over \hbar^2 }
P_x(t) - {2 B_{ac} g_n \mu_n \over \hbar}P_z(t),\nonumber\\ {d
P_y(t)\over dt} &=& {- 2 B_z^2 \epsilon \over \hbar^2 } P_y(t) ,
\nonumber\\ {d P_z(t)\over dt} &=& {2 B_{ac} g_n \mu_n \over \hbar
} P_x(t),
\end{eqnarray}
that can be solved to give
\begin{eqnarray}
P_x(t) &=& {\rm exp}[{- B_z^2 \epsilon t \over  \hbar^2}] \{ ({\rm cosh} ({\alpha t \over \hbar^2}) - {
B_z^2 \epsilon \over \alpha}  {\rm sinh}({\alpha t \over \hbar^2})) P_x(0) - {2
B_{ac} g_n \mu_n \hbar \over \alpha} {\rm sinh}({\alpha t \over \hbar^2}) P_z(0) \}
, \nonumber\\ 
\label{equation:exactsoln} 
P_y(t) &=& {\rm exp}[{- 2
B_z^2  \epsilon t \over \hbar^2}] P_y(0), \nonumber\\ 
P_z(t) &=&
{\rm exp}[{- B_z^2 \epsilon t \over  \hbar^2}] \{
({\rm cosh} ({\alpha t \over \hbar^2}) + { B_z^2 \epsilon \over \alpha}
{\rm sinh}({\alpha t \over \hbar^2})) P_z(0) + {2 B_{ac}
g_n \mu_n \hbar \over \alpha}{\rm sinh}({\alpha t \over \hbar^2}) P_x(0)
\}.\nonumber\\
\end{eqnarray}
Here we have defined $\alpha = \sqrt{ B_z^4 \epsilon^2 - 4
(B_{ac} g_n \mu_n\hbar)^2}$. In this case we see that all the
components of the polarization vector decay to zero, thus leaving
a totally mixed state regardless of the initial state. This is
known as a depolarizing process, not only is the phase coherence
lost, but the population probabilities become uniform and the
qubit is equally likely to be in the classical $|0\rangle$ or
$|1\rangle$ state. The action of the single qubit gate is to mix
the $P_x$ and $P_z$ components of the polarization vector. The
dephasing process causes the decay of the $P_x$ and $P_y$
components and these two processes combine to cause the
depolarization. The two processes define two time scales in the
system, the first $\tau_{op} = {\pi \hbar \over 4 B_{ac}
g_n \mu_n} $ is the time it takes to perform a typical single
qubit logic operation, a Hadamard gate. The time scale of the
dephasing process is given by $\tau_{dec} = {\hbar^2 \over 2 B_z^2
\epsilon}$. Obviously a functioning quantum computer requires
$\tau_{op} / \tau_{dec} << 1$. To zeroth order in this ratio 
we find that Eqs(\ref{equation:exactsoln}) give
\begin{eqnarray}
P_x(t) &=& {\rm exp}[{- B_z^2 \epsilon t \over \hbar^2}] \{ {\rm cos}({- 2 B_{ac}g_n \mu_n t \over \hbar})  P_x(0) +
{\rm sin}({-2 B_{ac}g_n \mu_n t \over \hbar}) P_z(0) \} , \nonumber\\
\label{equation:fstapprox}
P_y(t) &=& {\rm exp}[{- 2 B_z^2 \epsilon t \over \hbar^2}] P_y(0), \\
P_z(t) &=& {\rm exp}[{- B_z^2 \epsilon t \over \hbar^2}] \{ {\rm cos}({- 2 B_{ac}g_n \mu_n t \over \hbar}) P_z(0) -
{\rm sin}({-2 B_{ac} g_n \mu_n t \over \hbar}) P_x(0) \} . \nonumber
\end{eqnarray}
We can now calculate the fidelity of these noisy operations
\begin{eqnarray}
F &=& {\rm Tr} [ {\tilde \rho_{av}}(t) \rho (t) ] \nonumber\\ &=& {1
\over 2} ( 1 + {\rm exp}[{-2 B_z^2 \epsilon t \over \hbar^2}]
P_y(0)^2 + {\rm exp}[{-B_z^2 \epsilon t \over  \hbar^2}]
(P_x(0)^2+P_z(0)^2), \label{equation:sndapprox}
\end{eqnarray}
where $\rho (t)$ is the output density operator for a noiseless
operation. In this case the fidelity decays to
${1\over2}$ regardless of the input state, however the rate of decay is
dependent of the initial state. We see that the rate of
fidelity loss is not faster than is the case for a qubit in the
quantum register and in the worst case they are equal. The worst
case occurs, for $y$ rotations when the qubit is initially in a
$\sigma^y$ eigenstate, this means that the gate cannot rotate the
state out of the basis in which it decoheres.
\section{Tolerance to Noise}
\par To calculate the level of noise in the voltage signal that the quantum computer can tolerate in performing a
typical single particle operation, the Hadamard gate. There is
still some debate as to how much error a QC can tolerate and still
function usefully, even with error correcting codes. The estimates
range between an error probability of $\delta = 10^{-6} - 10^{-4}$
per qubit per operation \cite{Preskill,Aharonov}. For the purposes
of this calculation we will hedge our bets and use the limit
$\delta = 10^{-5}$. We use the operating parameters prescribed by
Kane \cite{Kane}; $B_z = 2T$ and $ B_{ac} = 0.001T $ and an A-gate
bias of $1V$, a value at the top of the bias range. We then find
that using the worst case fidelity function given by
eq(\ref{equation:worstfidel}) the error probability is given by
\begin{eqnarray}
\delta &=&  1-F \nonumber\\ &=&{1\over2}(1-{\rm exp}[{-
\tau_{op}\over \tau_{dec}}]),
\end{eqnarray}
which substituting in our limit for $\delta$ gives
\begin{equation}
{\tau_{op} \over \tau_{dec}} < 2 \times 10^{-5}.
\end{equation}
This limit justifies our zeroth order approximation in obtaining eqs(\ref{equation:fstapprox},\ref{equation:sndapprox}).
We now require
\begin{equation}
\epsilon < {B_{ac} g_n \mu_n \hbar  \over \pi B_z^2 }\times 4 \times
10^{-5}.
\end{equation}
Using eq(\ref{equation:epslam}) we find
\begin{equation}
\lambda = {B_{ac} g_n \mu_n \over \pi \eta^2 V_0^2 \hbar}\times 4
\times 10^{-5},
\end{equation}
and so we get a limit on the acceptable noise in the voltage signal. To implement a Hadamard gate requires a pulse of
duration $t = \tau_{op}$, thus  we find that the ratio of the rms fluctuations in the pulse area to the mean pulse area
is restricted by
\begin{equation}
{\Delta \Gamma(t_{Had})_{rms} \over {\bar \Gamma(t_{Had})}} < {B_{ac} \gamma'  \over \pi
\eta V_0 \hbar} \times 1.3 \times 10^{-2},
\end{equation}
which under the operating conditions of the Kane computer gives
\begin{equation}
{\Delta \Gamma(t_{Had})_{rms} \over {\bar \Gamma(t_{Had})}} <  1.4 \times 10^{-6}.
\end{equation}
\section{Conclusions}
\par We have found that the stochastic white noise process which causes dephasing of qubits in the quantum register
becomes a depolarizing process for qubits undergoing rotations in
the quantum computer. Under similar operating conditions the
fidelity of these qubit rotations decays more slowly for certain
highly coherent input states, than does the fidelity of the
register. We find that in order to satisfy currently accepted
limits of on error probability per qubit operation we must keep
the ratio of the rms fluctuations in the pulse area to the
intended pulse area ${\Delta \Gamma(t_{Had})_{rms} \over {\bar \Gamma(t_{Had})}} < 1.4
\times 10^{-6}$.
\section{Acknowledgments}
\par CJW would like to acknowledge the support of an Australian Postgraduate Award, a Melbourne University Postgraduate
Abroad Scholarship and the Max-Planck-Institut f\"ur Kernphysik. CJW would also like to acknowledge valuable discussions
with B.H.J.McKellar, and the theory group at the University of Queensland. LCLH wishes to acknowledge the support of the Alexander von Humboldt
foundation and the  Max-Planck-Institut f\"ur Kernphysik.

\end{document}